\documentclass{amsart}
\usepackage{amsmath, amsthm, amsfonts, amssymb}
\newtheorem{theorem}{Theorem}[section]
\newtheorem{lemma}[theorem]{Lemma}

\theoremstyle{definition}

\newtheorem{heuristic}[theorem]{Heuristic Assumption}

\newtheorem{obstruction}[theorem]{Obstruction}

\theoremstyle{remark}

\numberwithin{equation}{section}

\newcommand{\Z}{\mathbb Z}
\newcommand{\F}{\mathbb F}

\begin{document}

\title[On the relation generation method of Joux for discrete logarithms.]{On the relation generation method of Joux for computing discrete logarithms.}

\author{Ming-Deh Huang }
\address{Computer Science Department, University of Southern California, Los Angeles.}
\email{mdhuang@usc.edu }

\author{Anand Kumar Narayanan}
\address{Computer Science Department, University of Southern California, Los Angeles.}
\email{aknaraya@usc.edu}

\begin{abstract}
In \cite{joux}, Joux devised an algorithm to compute discrete logarithms between elements in a certain subset of the multiplicative group of an extension of the finite field $\mathbb{F}_{p^n}$ in time polynomial in $p$ and $n$. Shortly after, Barbulescu, Gaudry, Joux and Thome \cite{bgjt} proposed a descent algorithm that in $(p n)^{\mathcal{O}(\log n)}$ time projects an arbitrary element in $\mathbb{F}_{p^n}^\times$ as a product of powers of elements in the aforementioned subset. Together, these two algorithms yield a quasi-polynomial time algorithm for computing discrete logarithms in finite fields of small characteristic. The success of both the algorithms are reliant on heuristic assumptions. We identify obstructions that prevent certain heuristic assumptions they make from being true in general. Further, we describe methods to overcome these obstructions.  
\end{abstract}
\maketitle
\section{Introduction}\label{intro}
The discrete logarithm problem over the finite field $\F_{p^n}$ is given a generator $\eta$ of the multiplicative group $\F_{p^n}^\times$ and an element $\gamma \in \F_{p^n}^\times$ to find an integer $\log_\eta(\gamma)$ uniquely determined modulo $p^n-1$ such that $\gamma = \eta ^{\log_\eta(\gamma)}$. The discrete logarithm problem serves as an important cryptographic primitive. For instance, the security of the Diffie-Hellman \cite{dh} key exchange protocol and ElGamal's cryptosystem \cite{elg} are conditioned on the hardness of the discrete logarithm problem over finite fields.\\ \\
Let $L(\ell)$ denote $\exp( \mathcal{O}((\log(p^n)^\ell)(\log\log(p^n))^{1-\ell}))$. The index calculus method has been developed for the discrete logarithm problem over finite fields in a series of works \cite{adl}\cite{cop}\cite{gor}\cite{adl1}\cite{ah}\cite{jl}\cite{jlsv} establishing that the problem can be solved in $L(1/3)$ time which is subexponential in $\log(p^n)$.\\ \\
In recent breakthroughs, Gologlu, Granger, McGuire, Zumbragel \cite{gggz} and Joux \cite{joux} independently devised algorithms that assuming certain heuristics compute discrete logarithms in small characteristic finite fields faster than previously known. The authors of \cite{gggz} demonstrated their algorithm by computing discrete logarithms in $\F_{2^{1971}}$ which at the time of announcement was a record \cite{gggz1}. Joux's algorithm is the first to compute discrete logarithms in heuristic $L(1/4+o(1))$ time. This speed up allowed Joux \cite{joux2} to compute discrete logarithms in $\F_{2^{4080}}$. Gologlu, Granger, McGuire and Zumbragel \cite{gggz2} then extended the record to $\F_{2^{6120}}$. \\ \\
A remarkable feature shared by the algorithms in \cite{gggz} and \cite{joux} is that in their initial phases, they both consider a small set as the factorbase, one that is of size polynomial in the extension degree. Further, the factorbase consists of elements that can be represented as linear polynomials in $\zeta$. Here $\zeta$ is a root of a carefully chosen irreducible polynomial that they adjoin to construct the finite field. Assuming their relation generation algorithms succeed, discrete logarithms of the factor base elements can be determined up to a common constant multiple.\\ \\
In a recent further advancement Barbulescu, Gaudry, Joux and Thome \cite{bgjt} proposed an algorithm for computing discrete logarithms in small characteristic finite fields in quasi polynomial time. The finite field representation chosen in \cite{bgjt} is identical to the one in Joux's algorithm \cite{joux}. The descent phase in \cite{bgjt} expresses an arbitrary element in the multiplicative group of the finite field as a product of powers of elements in the factorbase in \cite{joux}. Thereby, the descent in \cite{bgjt} reduces the discrete logarithm computation over the finite field to computing discrete logarithms between elements in the factorbase which can be solved efficiently by the initial phase in Joux's algorithm \cite{joux}.\\ \\
The success of the initial phase of Joux's algorithm in determining the discrete logarithm between elements in the factorbase and the success of the descent in \cite{bgjt} are both reliant on heuristic assumptions. However there are obstructions that prevent these heuristics from being true in general as illustrated in \S~\ref{obstruction_section}. We propose modifications to the polynomial selection in Joux's algorithm in \S~\ref{poly_search} and in section \S~\ref{dlog_snf} design an algorithm for computing discrete logarithms in the factorbase. With the modified polynomial selection, if the relation generation is successful in generating a large enough lattice, the algorithm in \S~\ref{dlog_snf} succeeds. In \S~\ref{descent}, we discuss our proposed modification to the descent phase.
\subsection{Obstruction to Computing Discrete Logarithms in the Factorbase}\label{obstruction_section}
Joux's algorithm first proceeds by embedding $\F_{p^n}$ into an extension $\F_{q^{2n}}$ where $q$ is a power of $p$ such that $n \leq q$. 
The field $\F_{q^{2n}}$ is constructed as $\F_{q^2}[\zeta]$, where $\zeta$ is a root of an irreducible polynomial $g(x) \in \F_{q^2}[x]$ of degree $n$ that is of the following special form. Polynomials $h_0(x), h_1(x) \in \F_{q^2}[x]$ of low degree such that the factorization of $h(x) := h_1(x)x^q - h_0(x)$ over $\F_{q^2}[x]$ has an irreducible factor of degree $n$ are sought and one such irreducible factor of degree $n$ is picked as $g(x)$. The motivation behind choosing $g(x)$ in this manner is that the identity $h_1(x) x^q - h_0(x)=0 \mod g(x)$ is used by the relation generation algorithm to replace $x^q \mod g(x)$ with an expression consisting of the low degree polynomials $h_0(x)$ and $h_1(x)$ modulo $g(x)$.\\ \\ 
The set of monic linear polynomials in $\F_{q^2}[x]$ modulo $g(x)$ along with $h(x) \mod g(x)$ and a generator $\lambda$ of $\F_{q^2}^\times$ is taken as the factorbase $F$ and \cite[Thm. 8]{chung}\cite[Ques 1.1]{wan} guarantees that $\langle F \rangle \cong \F_{q^{2n}}^\times$. The relation generation phase collects multiplicative relations between the elements in $F$. Formally, an identity in $\langle F \rangle$ of the form $\prod_{\beta \in F} \beta^{e_\beta} = 1$ for integers $e_\beta$ is called as a relation and can be identified with the relation vector $(e_\beta)_{\beta \in F} \in \Z^{|F|}$ indexed by elements in $F$. Let $Rel$ denote the set of relation vectors collected and let $$\Gamma_R :=  \sum_{(e_\beta)_{\beta \in F} \in Rel} \Z\ (e_\beta)_{\beta \in F}$$  
denote the $\Z$-lattice generated by the collected relations.\\ \\
For a polynomial $f(x) \in \F_{q^2}[x]$, let $\F_f$ denote the ring $\F_{q^2}[x]/\left( f(x)\F_{q^2}[x]\right)$.\\ \\ 
To illustrate the obstruction, we restrict our attention to the case where $h(x)$ is square free. Let $h(x)= \prod_{i=0}^k g_i(x)$ be a factorization of $h(x)$ into distinct irreducible polynomials in $\F_{q^2}[x]$. Without loss of generality, let $g_0(x)=g(x)$.\\ \\
For a non constant polynomial $f(x) \in \F_{q^2}[x]$ dividing $h(x)$, let $\Gamma_f$ denote the relation lattice of the subgroup of $\F_{f}^\times$ corresponding to the generating set $$F_f: =  \{\lambda\} \cup \{h_1(x) \mod f(x)\} \cup \{x + \theta \mod f(x) , \theta \in \F_{q^2}\}.$$
That is, $$\Gamma_f = \left\{ (z_\beta)_{\beta \in F_f}  \in \Z^{|F|} | \prod_{\beta \in F_f}\beta^{z_\beta} =1 \right\}$$ and $$\Z^{|F|}/\Gamma_f \cong \langle F_f \rangle.$$
\begin{obstruction}
There is an obstruction if there is a large prime $\ell$ dividing $|\F_g^\times|$ such that the $\ell$-primary part of $\langle F_h \rangle$ is not cyclic.
\end{obstruction}
\ \\For the rest of the section, fix a large prime $\ell$ that divides $|\F_g^\times|$.\\ \\
If the $\ell$-primary part of $\langle F_h \rangle$ is not cyclic, then $\langle F_h \rangle/\langle F_h \rangle^\ell$ is not cyclic.\\ \\
If $\langle F_h \rangle/\langle F_h \rangle^\ell$ is not cyclic, then discrete logarithms of elements in the factorbase cannot be determined from the relation lattice $\Gamma_R$ as stated in \cite{bgjt}[Prop 2, part 2] for the following reason.\\ \\
While the intent was to collect relations in $\F_g^\times$, the relations generated are satisfied in $\F_h^\times$. In fact, every relation generated is satisfied in $\langle F_h \rangle$. The obstruction is structural and intrinsic to the polynomial $h(x)$ and applies to every relation generation algorithm that is restricted to relations in $\F_h^\times$.\\ \\
Since $\Gamma_R$ is a sublattice of $\Gamma_h$ there is an inclusion $$\langle F_h \rangle \hookrightarrow \Z^{|F|}/\Gamma_R.$$
Thus if $\langle F_h \rangle/\langle F_h \rangle^\ell$ is not cyclic, then $(\Z^{|F|}/\Gamma_R)/\ell(\Z^{|F|}/\Gamma_R)$ is not cyclic.\\ \\
For $\beta \in F$, let $\theta_\beta \in \Z$ (uniquely determined modulo $|\F_g^\times|$) denote the discrete logarithm in $\F_g^\times$ of $\beta$ with respect to a chosen generator of $\F_g^\times$.\\ \\
If $\forall \beta \in F, \theta_\beta = 0 \mod \ell$, then $\langle F \rangle$ is a subgroup of $\left(\F_g^\times\right)^\ell$ which contradicts the fact that $\langle F \rangle = \F_g^\times$. Thus there exists an $\alpha \in F$ such that $\theta_\alpha \neq 0 \mod \ell$.\\ \\
Consider the problem of determining $\theta_\beta/\theta_\alpha \mod \ell$ for all $\beta \in F$. This is at least as easy as determining $\theta_\beta \mod \ell$ for all $\beta \in F$.\\ \\
To infer $\theta_\beta/\theta_\alpha \mod \ell$ from the collected set of relations $Rel$, it is necessary that the linear system \begin{equation}\label{linear_system} \left\{ e_\alpha + \sum_{\beta \in F -\{\alpha\}}{e_\beta (\theta_\beta/\theta_\alpha)}  = 0 \mod \ell,\ (e_\beta)_{\beta \in F} \in Rel\right\}\end{equation} over $\Z/\ell\Z$ is of rank $|F|-1$.\\ \\
Let $R(\ell)$ denote the $\Z/\ell\Z$ module generated by $Rel$. If the linear system \ref{linear_system} is of rank $|F|-1$ over $\Z/\ell\Z$, then $(\Z/\ell\Z)^{|F|}/R(\ell)$ is cyclic.\\ \\ Since $(\Z/\ell\Z)^{|F|}/R(\ell) \cong (\Z^{|F|}/\Gamma_R)/\ell(\Z^{|F|}/\Gamma_R)$, if the rank of the linear system \ref{linear_system} over $\Z/\ell\Z$ is $|F|-1$, then $(\Z^{|F|}/\Gamma_R)/\ell(\Z^{|F|}/\Gamma_R)$ is cyclic.\\ \\
Thus to infer $\forall \beta \in F, \theta(\beta) \mod \ell$, it is necessary that $(\Z^{|F|}/\Gamma_R)/\ell(\Z^{|F|}/\Gamma_R)$ is cyclic. If the $\ell$-primary part of $\langle F_h \rangle$ is not cyclic, then $(\Z^{|F|}/\Gamma_R)/\ell(\Z^{|F|}/\Gamma_R)$ is not cyclic thereby obstructing the discrete logarithm computation. $\square$.\\ \\
We next describe choices of $h(x)$ for which an obstruction is likely to occur.\\ \\
For all $i \in \{0,1,\ldots, k\}$, since $\deg(g_i) \leq q$, \cite[Thm. 8]{chung}\cite[Ques 1.1]{wan} implies that $\Z^{|F|}/\Gamma_{g_i} \cong \langle F_{g_i} \rangle \cong \F_{g_i}^\times$. Further, since $\Gamma_h \subseteq \Gamma_{g_i}$ there is a natural surjection $\Z^{|F|}/\Gamma_h \twoheadrightarrow \Z^{|F|}/\Gamma_{g_i}$ which since $\Z^{|F|}/\Gamma_{g_i} \cong \F_{g_i}^\times$ implies that there is a surjection $$\langle F_h \rangle  \twoheadrightarrow \F_{g_i}^\times.$$ Since the orders of the groups in $\{\F_{g_i}^\times\ |\ 0 \leq i \leq k\}$ are not pairwise relatively prime, $\langle F_h \rangle $ is not likely to be cyclic.\\ \\
The orders of the groups in $\{\F_{g_i}^\times\ |\ 0 \leq i \leq k\}$ are not relatively prime since every $\F_{g_i}^\times$ contains $\F_{q^2}^\times$ as a subgroup. This however is not a concern if $\langle F_h\rangle^{q^2-1}$ is cyclic since in this case we can compute discrete logarithms in $\langle F_h\rangle^{q^2-1}$ and later account for the $\F_{q^2}^\times$ part.\\ \\ 
The concern is when $h(x)$ has a factor $g_i(x)$ other than $g(x)$ such that $|\F_g^\times|$ and $|\F_{g_i}^\times|$ share a large prime factor $\ell$ as it is then likely that the $\ell$-primary part of $\langle F_h\rangle$ is not cyclic.\\ \\
For instance when the degrees of $g_i(x)$ and $g(x)$ share a large enough factor, the existence of a large prime factor dividing both $|\F_g^\times|$ and $|\F_{g_i}^\times|$ is all but certain.\\ \\
A particularly acute and illustrative case is when $h(x)$ has an irreducible factor $g_i(x)$ other than $g(x)$ of degree $n$. In this case, every relation generated holds modulo both $g(x)$ and $g_i(x)$. Further, $\langle F_h\rangle$ surjects to both $(\F_g^\times)/(\F_g^\times)^\ell$ and $(\F_{g_i}^\times)/(\F_{g_i}^\times)^\ell$. Since $\Gamma_g$ and $\Gamma_{g_i}$ are not likely to be identical, it is likely that the $\ell$-primary part of $\langle F_h\rangle$ is not cyclic. In this case, to compute discrete logarithms in $\F_g^\times$, relations that hold modulo both $g(x)$ and $g_i(x)$ do not suffice and we require to break the symmetry between $g(x)$ and $g_i(x)$ by finding relations that hold modulo $g(x)$ but not modulo $g_i(x)$.\\ \\
To overcome the obstruction, in \S~\ref{poly_search} we propose imposing further restrictions on $h(x)$ that ensure that for every large prime $\ell$ dividing $|\F_g^\times|$, the $\ell$-primary part of $\langle F_h\rangle$ is cyclic. For instance, we insist that $\gcd(|\F_{h/g}^\times|, |\F_g^\times|)$ is smooth. The restrictive polynomial selection ensures that if the relation generation algorithm succeeds in determining $\Gamma_h$ or a large enough sublattice of $\Gamma_h$, then we can project the elements in the factorbase efficiently to a certain large order cyclic subgroup of $\F_g^\times$ and compute discrete logarithms there. The complement under direct product of this large order subgroup in $\F_g^\times$ has smooth order and hence can be dealt with the Pohlig-Hellman algorithm.\\ \\
In \S~\ref{dlog_snf} we describe an algorithm for computing discrete logarithms that involves computing certain invariant factor decompositions. As a consequence we have the following Theorem \ref{discrete_log_theorem_intro} which states that given a polynomial $h(x)$ as in \S~\ref{poly_search}, if the relation lattice $\Gamma_R$ generated is the entire relation lattice $\Gamma_h$ of the factorbase modulo $h(x)$, then the discrete logarithm between elements in the factorbase can be computed.
\begin{theorem}\label{discrete_log_theorem_intro}
If $\Gamma_R = \Gamma_h$, then \\ \\
(1) a generator $\mu$ of $\F_g^\times$ can be found in $q^{\mathcal{O}(1)}$-time.\\ \\
(2) $\forall \beta \in F$, a $\theta_\beta \in \Z$ such that $\mu^{\theta_\beta} = \beta$ can be found the in $q^{\mathcal{O}(1)}$-time. 
\end{theorem}
\noindent Since $q$ is bounded by a polynomial in $p$ and $n$, $q^{\mathcal{O}(1)}$ is bounded by a polynomial in $p$ and $n$ and the algorithm is efficient in small characteristic.\\ \\
 In \S~\ref{alternate}, we present an alternate algorithm that computes discrete logarithms by performing a sequence of row operations on the relation matrix.\\ \\
Another way to avoid the obstruction is to require that $|\left(\F_h^\times\right)^{s}|$ is square free for some smooth number $s$. In this case, we are assured the $\left(\F_h^\times\right)^s$ is cyclic (and thus $\langle F_h\rangle^s$ is cyclic). The $s$-smooth part can be dealt with using the Pohlig-Hellman algorithm. The requirement that we insist on is less restrictive.\\ \\
We note that the obstruction is easy to resolve in the Kummer case, that is when $n=q-1$. When $n=q-1$, $h(x)$ is chosen as $x^q-\lambda x$ and $g(x)$ as $x^{q-1}-\lambda$ where $\lambda$ is a generator of $\F_{q^2}^\times$. In this case, $h(x) = xg(x)$ and thus $(\F_h^\times)^{q^2-1}$ is cyclic. Further, the relation $x^{q-1} = \lambda  \mod g(x)$ can be added to the relation matrix and this allows the inclusion of $x \mod g(x)$ in the factorbase.\\ \\
We became aware of this obstruction and proposed the modified polynomial selection in \cite{primitive} while using Joux's algorithm to efficiently find primitive elements in finite fields of small characteristic. This obstruction was described independently by Cheng, Wan and Zhang \cite{cwz} who suggested a different modification.
\subsection{Implications on the Descent }
The descent proposed in \cite{bgjt} involves generating certain multiplicative relations in $\F_h^\times$ and is based on techniques related to Joux's relation generation algorithm \cite{joux} to collect relations between the elements in the factorbase. The fact that the relations generated hold modulo $h(x)$ is a concern for the descent algorithm as well.\\ \\
Each step in the descent starts with a given polynomial $P(x) \in \F_{q^2}[x]$ of degree $w < n$ and multiplicative relations modulo $h(x)$ between $\{P(x)+\beta, \beta \in \F_{q^2}\}$ and polynomials of degree at most $w/2$ are generated. If sufficiently many relations are generated, then we attempt to express $P(x)$ modulo $h(x)$ as a product of powers of polynomials of degree at most $w/2$. Then descent steps starting from the polynomials of degree at most $w/2$ that appear is performed recursively until we are left with linear polynomials.\\ \\
Let $U$ denote the set of monic irreducible polynomials in $\F_{q^2}[x]$ of degree at most $w/2$ and $G_U$ denote the subgroup of $\F_h^\times$ generated by the images of elements in $U$. Consider the natural surjection $$\psi : \F_h^\times \longrightarrow (\F_h^\times)/(\F_h^\times)^\ell.$$ 
Since every relation generated holds in $\F_h^\times$, for the descent step to succeed it is necessary that there exists $e_u \in \Z$ such that the images of $P(x)$ and $\prod_{u \in U}u^{e_u}$ in $(\F_h^\times)/(\F_h^\times)^\ell$ are identical. That is, for the descent step to succeed $\psi(P(x) \mod h(x))$ needs to be in $\psi(G_U)$.\\ \\
If $\psi$ is not surjective, then $$\frac{\left|\psi(G_U)\right|}{\left|(\F_h^\times)/(\F_h^\times)^\ell \right| } \leq \frac{1}{\ell}$$
and heuristically it is likely that the descent from a polynomial of degree $w$ will fail with the possible exception of at most $1/\ell$ fraction of polynomials of degree $w$.\\ \\
If $(\F_h^\times)/(\F_h^\times)^\ell$ is cyclic, then $\psi$ is surjective since the set of monic linear polynomials is contained in $U$ and by \cite[Thm. 8]{chung}\cite[Ques 1.1]{wan} the images of linear polynomials generate $\F_{g_i}^\times$ for every $g_i(x)$ dividing $h(x)$.\\ \\
If $(\F_h^\times)/(\F_h^\times)^\ell$ is not cyclic, then $\psi$ may not be surjective. This is a concern, especially when $w$ is small.\\ \\
Consider the descent from quadratic polynomials, that is $w=2$. In this case, $G_U = \langle F_h \rangle$ (for ease of exposition ignoring the fact that $h_1(x) \mod h(x)$ is in $F_h$). Suppose there is a large prime $\ell$ dividing $|\F_g^\times|$ and $|\F_{g_i}^\times|$ for some $g_i(x)$ dividing $h(x)/g(x)$.\\ \\ 
In this case, $(\F_h^\times)/(\F_h^\times)^\ell$ is not cyclic. If $\psi$ is not surjective, then the descent from a quadratic polynomial is likely to fail. If $\psi$ is surjective, then $(\langle F_h\rangle)/(\langle F_h\rangle)^\ell (= (\F_h^\times)/(\F_h^\times)^\ell)$ is not cyclic and the computation of discrete logarithms in the factorbase fails due to obstruction.\\ \\
It is thus necessary that for every factor $g_i(x)$ of $h(x)/g(x)$, $|\F_g^\times|$ and $|\F_{g_i}^\times|$ do not share a large prime factor.\\ \\
In addition Cheng, Wan and Zhang \cite{cwz} described potential traps to the descent algorithm in \cite{bgjt} that prevent the descent from succeeding and suggested a trap avoiding descent. In the trap avoiding descent of \cite{cwz}, certain relations involving the factors of $h(x)$ are identified as traps and are not used when encountered. This trap avoidance thus comes at the cost of discarding relations.\\ \\  
In \S~\ref{descent}, we observe that some of the relations dropped in fear of traps could be salvaged. Further, these salvaged relations serve in further breaking the symmetry between $g(x)$ and the other irreducible factors of $h(x)$.
\section{Discrete Logarithms in the Factorbase}\label{discrete_log_factorbase}
\subsection{Polynomial Selection}\label{poly_sel}
Recall from the introduction section that for a polynomial $f(x) \in \F_{q^2}[x]$, $\F_f$ denotes the ring $\F_{q^2}[x]/\left( f(x)\F_{q^2}[x]\right)$. For this section, fix a polynomial $h(x) \in \F_{q^2}[x]$ that satisfies the following three conditions. 
\begin{enumerate}
\item $h(x)$ has an irreducible factor of degree $m$ (call it $g(x)$).
\item The square of $g(x)$ does not divide $h(x)$.
\item $\gcd(\left|\F_{h/g}^\times \right|, q^{2m}-1)$ is $q^{2C}$-smooth.
\end{enumerate}
Let $$h(x) = g(x) \prod_{i=1}^k g_i(x)^{a_i}$$ be a factorization where $g_i(x)$ are distinct irreducible polynomials in $\F_{q^2}[x]$.\\ \\
We are interested in computing discrete logarithms in $\F_g^\times \cong \F_{q^{2m}}^\times$.\\ \\
Fix a positive integer $C$ that defines a smoothness bound. We say that an integer is $q^{2C}$-smooth if and only if all its prime factors are at most $q^{2C}$.
\subsection{Computing Discrete Logarithms in the Factorbase: Algorithm I}\label{dlog_snf}
Let $D \subset \F_{q^2}[x]$ be a finite set of polynomials such that the set $F :=\{d(x) \mod g(x) | d(x) \in D\} \subset \F_g^\times$ of polynomials in $D$ modulo $g(x)$ satisfies $\langle F \rangle = \F_g^\times$.\\ \\
An identity in $\langle F \rangle \cong \F_{q^{2m}}^\times$ of the form $\prod_{\beta \in F} \beta^{e_\beta} = 1$ for integers $e_\beta$ is called as a relation and it can be identified with the relation vector $(e_\beta, \beta \in F) \in \Z^{|F|}$ indexed by elements in $F$. Let $R$ be the $N$ by $|F|$ matrix consisting of the relation vectors found by a relation generation algorithm as rows.\\ \\
For a non constant polynomial $f(x) \in \F_{q^2}[x]$ dividing $h(x)$, $\Gamma_f$ denotes the relation lattice of the subgroup of $\F_{f}^\times$ corresponding to the generating set $$F_f: =  \{d(x) \mod f(x)\ |\ d(x) \in D\}.$$
That is, $$\Gamma_f = \left\{ (z_\beta)_{\beta \in F_f}  \in \Z^{|F|} | \prod_{\beta \in F_f}\beta^{z_\beta} =1 \right\}.$$\\
Let $\Gamma_R$ be the $\Z$-lattice generated by the collected relations, that is by the rows of the matrix $R$. Since $\Gamma_R$ is contained in $\Gamma_h$ which is in turn contained in $\Gamma_g$ and we have the natural surjection $$ \Z^{|F|}/\Gamma_R \twoheadrightarrow \Z^{|F|} /\Gamma_g.$$
Since $\Z^{|F|}/\Gamma_g \cong \F_g^\times$, we have the natural surjection $\varphi : \Z^{|F|}/\Gamma_R \twoheadrightarrow \F_g^\times$.\\ \\
The Smith normal form of $R$ gives the decomposition of $\Z^{|F|}/\Gamma_R$ into its invariant factors $$\Z^{|F|}/\Gamma_R = \langle e(1) \rangle \oplus \langle e(2) \rangle \oplus \ldots \oplus \langle e(|F|) \rangle  \cong  \Z/d_1\Z \oplus \Z/d_2\Z \oplus \ldots \oplus \Z/d_{|F|}\Z $$
where for $1\leq i \leq |F|$, $e(i) \in \Z^{|F|}$ denotes a relation vector and $d_i$ the order of $e(i)$ in $\Z^{|F|}/\Gamma_R$ and for $1 \leq i <|F|$, $d_i \mid d_{i+1}$.\\ \\
For $1 \leq i <|F|$, let $\pi_{i}$ denote $\varphi(e(i)) = \prod_{\beta \in F} \beta^{e(i)_\beta}$.\\ \\
We next prove a lemma which states a condition on $\Z^{|F|}/\Gamma_R$ that guarantees that our relation generation step has collected enough enough relations to determine a certain large order cyclic subgroup in $\F_g^\times$. If the relations determine the aforementioned cyclic subgroup, then the discrete logarithm between elements in the factorbase can be determined.
\begin{lemma}\label{sufficient} If $\gcd(d_{|F|-1},q^{2m}-1)$ is $q^{2C}$-smooth, then there exists a $q^{2C}$-smooth number $B$ such that the order of  $\varphi(e(|F|))$ in $\F_g^\times$ is divisible by $\frac{q^{2m}-1}{B}$.
\end{lemma}
Assume $\gcd(d_{|F|-1},q^{2m}-1)$ is $q^{2C}$-smooth. From the Smith normal form, we have the invariant factor decomposition $$\Z^{|F|}/\Gamma_R = \bigoplus_{j=1}^{|F|} \langle e(j) \rangle$$ where $d_j$ is the order of $e(j)$ in $\Z^{|F|}/\Gamma_R$.\\ \\
Since $\left|\varphi\left(\bigoplus_{j=1}^{|F|-1} \langle e(j) \rangle\right) \right| = \prod_{j=1}^{|F|-1} \left|\varphi\left(\langle e(j) \rangle\right)\right|$ divides $\prod_{j=1}^{|F|-1} d_j$ and $d_j \mid d_{j+1}$ for $1 \leq j <|F|-1$, it follows that $\gcd\left(\left|\varphi\left(\bigoplus_{j=1}^{|F|-1} \langle e(j) \rangle\right) \right|,q^{2m}-1\right)$ is $q^{2C}$-smooth. \\ \\
Since $\varphi(\Z^{|F|}/\Gamma_R) = \F_g^\times$ and $\F_g^\times$ is cyclic of order $q^{2m}-1$, there exists a $q^{2C}$-smooth number $B$ such that the order of  $\varphi(e(|F|))$ in $\F_g^\times$ is divisible by $\frac{q^{2m}-1}{B}$. $\square$\\ \\
We next show if the relation generation is successful in computing the relation lattice of $\Gamma_h$ in its entirety, then the condition stated in lemma \ref{sufficient} is satisfied.
\begin{lemma}\label{lattice} If $\Gamma_R = \Gamma_h,$ then $\gcd\left(d_{|F|-1},q^{2m}-1\right)$ is $q^{2C}$-smooth.
\end{lemma}
Let $v$ denote the largest factor of $q^{2m}-1$ that is $q^{2C}$-smooth and let $L = (q^{2m}-1)/v$. Since
$$h(x) = g(x) \prod_{i=1}^k g_i(x)^{a_i}$$ where $g_i(x)$ are distinct irreducible polynomials in $\F_{q^2}[x]$, Chinese remainder theorem over $\F_{q^2}[x]$ implies 
$$\F_h^\times \cong \F_g^\times \times \prod_{i=1}^k \F_{g_i^{a_i}}^\times.$$
Let $\langle F_h \rangle$ denote the subgroup of $\F_h^\times$ generated by $F_h$. We have the inclusion $$\psi: \langle F_h \rangle \hookrightarrow \F_g^\times \times \prod_{i=1}^k \F_{g_i^{a_i}}^\times$$ $$\ \ \ \ \  \alpha \longmapsto \alpha_g \prod_{i} \alpha_{g_i} $$
Since the projection from $\langle F_h \rangle$ to $\F_g^\times$ is surjective, there exists a $\beta \in \langle F_h \rangle$ whose projection $\beta_g$ in $\F_g^\times$ is of order $q^{2m}-1$.\\ \\
The order of $\beta \in \langle F_h\rangle$ is divisible by the order of its projection $\beta_g \in \F_g^\times$. Hence $\langle F_h\rangle$ has an element of order $q^{2m}-1$ which implies that we have an inclusion $$\Z/L\Z \hookrightarrow \langle F_h \rangle$$ and hence $L$ divides $\left| \langle F_h\rangle\right|$.\\ \\
Since $\langle F_h \rangle \hookrightarrow \F_g^\times \times \prod_{i=1}^k \F_{g_i^{a_i}}^\times$, $\left| \langle F_h\rangle\right|$ divides $(q^{2m}-1)\left| \F_{h/g}^\times\right|$.\\ \\ 
Since $\gcd(\left| \F_{h/g}^\times\right|,q^{2m}-1)$ is $q^{2C}$-smooth, it follows that there exists integers $w,y$ such that $w$ is $q^{2C}$-smooth, $\gcd(L,y) = 1$ and $\left| \langle F_h\rangle\right| = L w y$.\\ \\
For every prime $\ell$ dividing $L$, the $\ell$-primary component of $\langle F_h\rangle$ is cyclic since $\Z/L\Z \hookrightarrow \langle F_h \rangle$ and $\left| \langle F_h\rangle\right|$ is $L$ times a factor relatively prime to $L$. Hence in the Smith normal form of $\langle F_h \rangle$, for every prime $\ell$ dividing $L$, the $\ell$-primary component of $\langle F_h\rangle$ is contained in the largest invariant factor. In particular, the largest invariant factor has order divisible by $L$.\\ \\
Since $\left| \langle F_h \rangle\right| = L w y$, it follows that the second largest invariant factor of $\langle F_h \rangle$ has order dividing $wy$. Since $w$ is $q^{2C}$-smooth and $\gcd(L,y)=1$, $\gcd(wy,q^{2m}-1)$ is $q^{2C}$-smooth.\\ \\
If $\Gamma_R = \Gamma_h$, then $\Z^{|F|}/\Gamma_R \cong \langle F_h \rangle$ and the order $d_{|F|-1}$ of the second largest invariant factor of $\Z^{|F|}/\Gamma_R$ divides $wy$. Thus $\gcd(d_{|F|-1},q^{2m}-1)$ is $q^{2C}$-smooth. $\square$\\ \\
The next lemma shows that if $\gcd(d_{|F|-1},q^{2m}-1)$ is $q^{2C}$-smooth, then given two elements in $\F_g^\times$, each expressed as a product of elements in the factorbase $F$, we can efficiently decide if the first element is in the subgroup generated by the other and if so compute the discrete logarithm of the first element with respect to the second element as the base.
\begin{lemma}\label{discrete_log}
If $\gcd(d_{|F|-1},q^{2m}-1)$ is $q^{2C}$-smooth, then given $(a_\beta)_{\beta \in F} \in \Z^{|F|}$ and $(b_\beta)_{\beta \in F} \in \Z^{|F|}$, in $q^{\mathcal{O}(1)}$ time we can decide if $\prod_{\beta \in F} \beta^{a_\beta} \in \F_g^\times$ is in the subgroup generated by $\prod_{\beta \in F} \beta^{b_\beta} \in \F_g^\times$ and if so find an integer $j$ such that $\prod_{\beta \in F} \beta^{a_\beta} = \left(\prod_{\beta \in F} \beta^{b_\beta}\right)^j.$
\end{lemma}
Recall that we denoted the largest factor of $q^{2m}-1$ that is $q^{2C}$-smooth by $v$ and $(q^{2m}-1)/v$ by $L$.\\ \\
Let $\F_g^\times[L]$ denote $\{\beta \in \F_g^\times| \beta^L = 1\}$ and $\F_g^\times[v]$ denote $\{\beta \in \F_g^\times| \beta^v = 1\}$. Since $L$ and $v$ are relatively prime, $$\F_g^\times = \F_g^\times[v] \times \F_g^\times[L].$$
We can project from $\F_g^\times$ to $\F_g^\times[v]$ by taking $L^{th}$ powers. Since the order of $\F_g^\times[v]$ is $q^{2C}$-smooth, the discrete logarithm problem in $\F_g^\times[v]$ can be solved in $q^{\mathcal{O}(1)}$ time using the Pohlig-Hellman algorithm \cite{ph}.\\ \\
All that remains is to address the discrete logarithm computation in $\F_g^\times[L]$.\\ \\ 
Assume $\gcd(d_{|F|-1},q^{2m}-1)$ is $q^{2C}$-smooth. Let $$ \vartheta : \Z^{|F|}/\Gamma_R \longrightarrow \langle e(|F|) \rangle $$
denote the projection from $\Z^{|F|}/\Gamma_R$ to its largest invariant factor.\\ \\
Given a $\Z^{|F|}$ representative of an element $\kappa \in \Z^{|F|}/\Gamma_R$, the Smith normal form of $R$ allows us to efficiently compute an integer $\theta(\kappa)$ such that $\vartheta(\kappa) = e(|F|)^{\theta(\kappa)}$.\\ \\
Under the surjection $\varphi$, we have $$\Z^{|F|}/\Gamma_R \xrightarrow{\ \ \vartheta  \ \ } \langle e(|F|)\rangle \xrightarrow{\ \ \ \  \varphi\ \ \ \  } \F_{g}^\times$$
$$\ \ \ \ \ \ \ \ \ \ \ \ \ \ \ \ \ \ \ \ \kappa \longmapsto e(|F|)^{\theta(\kappa)} \longmapsto \left(\varphi(e(|F|))\right)^{\theta(\kappa)}.$$
From Lemma \ref{sufficient}, there exists a $q^{2C}$-smooth number $B$ such that the order of $\varphi(e(|F|))$ in $\F_g^\times$ is divisible by $\frac{q^{2m}-1}{B}$. Thus, the order of $\varphi(e(|F|))$ in $\F_g^\times$ is divisible by $L$.\\ \\
Let $\bar{\varphi}$ denote $\varphi$ composed with the projection from $\F_g^\times$ to $\F_g^\times[L]$. Then we have $$\Z^{|F|}/\Gamma_R \xrightarrow{\ \ \vartheta  \ \ } \langle e(|F|)\rangle \xrightarrow{\ \ \ \  \bar{\varphi}\ \ \ \  } \F_{g}^\times[L]$$
$$\ \ \ \ \ \ \ \ \ \ \ \ \ \ \ \ \ \ \kappa \longmapsto e(|F|)^{\theta(\kappa)} \longmapsto \left(\bar{\varphi}(e(|F|))\right)^{\theta(\kappa)}.$$
Let the images of $(a_\beta)_{\beta \in F}$ and $(b_\beta)_{\beta \in F}$ in $\Z^{|F|}/\Gamma_R$ be $\kappa_1$ and $\kappa_2$ respectively. The images of $\prod_{\beta \in F} \beta^{a_\beta}$ and $\prod_{\beta \in F} \beta^{b_\beta}$ in $\F_g^\times[L]$ are $\left(\bar{\varphi}(e(|F|))\right)^{\theta(\kappa_1)}$ and $\left(\bar{\varphi}(e(|F|))\right)^{\theta(\kappa_2)}$.\\ \\
Since $L$ divides the order of $\varphi(e(|F|))$, $\langle \bar{\varphi}(e(|F|))\rangle = \F_g^\times[L]$. Thus, the image of $\prod_{\beta \in F} \beta^{a_\beta}$ in $\F_g^\times[L]$ is in the subgroup generated by the image of $\prod_{\beta \in F} \beta^{b_\beta}$ in $\F_g^\times$ if and only if there exists an integer $j$ such that $$\theta(\kappa_1) = j \theta(\kappa_2) \mod L.$$
If such an $j$ exists, then $j \mod L$ is the discrete logarithm of the image of $\prod_{\beta \in F} \beta^{a_\beta}$ in $\F_g^\times$ with respect to the image of $\prod_{\beta \in F} \beta^{b_\beta}$ in $\F_g^\times$ as the base.\\ \\
We can decide if such an $i$ exists and if so find one using the extended Euclidean algorithm efficiently. $\square$.\\ \\
Lemmas \ref{sufficient}, \ref{lattice} and \ref{discrete_log} imply Theorem \ref{discrete_log_theorem}.
\begin{theorem}\label{discrete_log_theorem}
If $\Gamma_R = \Gamma_h$ then given $(a_\beta)_{\beta \in F} \in \Z^{|F|}$ and $(b_\beta)_{\beta \in F} \in \Z^{|F|}$, in time polynomial in $q$ and $|R|$ we can decide if $\prod_{\beta \in F} \beta^{a_\beta} \in \F_g^\times$ is in the subgroup generated by $\prod_{\beta \in F} \beta^{b_\beta} \in \F_g^\times$ and if so find an integer $j$ such that $\prod_{\beta \in F} \beta^{a_\beta} = \left(\prod_{\beta \in F} \beta^{b_\beta}\right)^j.$
\end{theorem}
\noindent If $\Gamma_R = \Gamma_h$, then $\Gamma_R$ can be used to find a generator of $\F_g^\times$ in $q^{\mathcal{O}(1)}$-time \cite{primitive}. Further, the number of the relations collected $|R|$ (see \S~\ref{relation}) in Joux's algorithm is bounded by a polynomial in $q$ thereby implying Theorem \ref{discrete_log_theorem_intro}.
\subsection{Computing Discrete Logarithms in Factorbase:  Algorithm II}\label{alternate}
\ \\ \\ We keep the notation from \S~\ref{poly_sel} and \S~\ref{dlog_snf}. Recall in particular that we denoted the largest factor of $q^{2m}-1$ that is $q^{2C}$-smooth by $v$ and $(q^{2m}-1)/v$ by $L$. In this section, we show that if for all $\ell$ dividing $L$, the $\Z/\ell\Z$ rank of the relation matrix $R_\ell$ is $|F|-1$, then we can efficiently find a generator for $\F_g^\times$
and find the discrete logarithm of an element in the factorbase with respect to the computed generator. By the end of this section, we will see that assuming $\Z^{|F|}/\Gamma_R$ is finite, the condition that for all $\ell$ dividing $L$, the $\Z/\ell\Z$ rank of the relation matrix $R_\ell$ is $|F|-1$ is equivalent to $\gcd(d_{|F|-1},q^{2m}-1)$ being $q^{2C}$-smooth.\\ \\
As in \S~\ref{dlog_snf}, the discrete logarithm computation in the smooth component of $\F_g^\times$ can be computed using the Pohlig-Hellman algorithm and we can restrict our attention to computing in the $L$-torsion $\F_g^\times[L]$. Let $\ell$ be a prime dividing $L$.\\ \\
Taking $\Z/\ell\Z$ tensor products of the exact sequence $$0 \longrightarrow  \Gamma_R  \longrightarrow \Z^{|F|} \longrightarrow \Z^{|F|}/\Gamma_R \longrightarrow 0$$
induces the sequence \\
$$  \Gamma_R \otimes\Z/\ell\Z \longrightarrow \Z^{|F|} \otimes \Z/\ell\Z \longrightarrow (\Z^{|F|}/\Gamma_R)\otimes \Z/\ell\Z \longrightarrow 0.$$\\
which is exact due to the right exactness of tensoring.\\ \\
Thus we have the isomorphism $$(\Z^{|F|} \otimes \Z/\ell\Z)/(\Gamma_R \otimes\Z/\ell\Z) \cong (\Z^{|F|}/\Gamma_R)/\ell(\Z^{|F|}/\Gamma_R).$$
Both $(\Z^{|F|} \otimes \Z/\ell\Z)$ and $(\Gamma_R \otimes\Z/\ell\Z)$ are $\Z/\ell\Z$ vector spaces and $(\Z^{|F|} \otimes \Z/\ell\Z)$ is $|F|$ dimensional over $\Z/\ell\Z$.
Let $R_\ell$ denote the matrix $R$ with the entries reduced modulo $\ell$. The $\Z/\ell\Z$ rank of $(\Gamma_R \otimes\Z/\ell\Z)$ equals the $\Z/\ell\Z$ rank of $R_\ell$ and we have the following characterization.\\
\begin{enumerate}
\item $(\Z^{|F|}/\Gamma_R)/\ell(\Z^{|F|}/\Gamma_R)$ is trivial $\Leftrightarrow$ the $\Z/\ell\Z$ rank of $R_\ell$ is $|F|$.
\item $(\Z^{|F|}/\Gamma_R)/\ell(\Z^{|F|}/\Gamma_R)$ is cyclic and non-trivial $\Leftrightarrow$ the $\Z/\ell\Z$ rank of $R_\ell$ is $|F|-1$.
\item $(\Z^{|F|}/\Gamma_R)/\ell(\Z^{|F|}/\Gamma_R)$ is not cyclic $\Leftrightarrow$ the $\Z/\ell\Z$ rank of $R_\ell$ less than $|F|-1$.
\end{enumerate}
By our choice of $h(x)$, $\ell$ does not divide $|\F_{h/g}^\times|$ and it follows that $$\F_h^\times/(\F_h^\times)^\ell \cong \F_g^\times/(\F_g^\times)^\ell.$$
Since the image of the elements of $F_h$ in $\F_g^\times$ generate $\F_g^\times$, there is a surjection $$\phi_\ell: (\Z^{|F|}/\Gamma_R)/\ell(\Z^{|F|}/\Gamma_R) \twoheadrightarrow \F_h^\times/(\F_h^\times)^\ell \cong \F_g^\times/(\F_g^\times)^\ell.$$
Thus $(\Z^{|F|}/\Gamma_R)/\ell(\Z^{|F|}/\Gamma_R)$ is not trivial and the $\Z/\ell\Z$ rank of $R$ is at most $|F|-1$.\\ \\
If the $\Z/\ell\Z$ rank of $R_\ell$ is $|F|-1$, then $(\Z^{|F|}/\Gamma_R)/\ell(\Z^{|F|}/\Gamma_R)$ is cyclic and the discrete logarithm in $\F_h^\times/(\F_h^\times)^\ell$ of an element in the factorbase is determined by $R_\ell$.\\ \\
However, unless we factor $L$, we do not know $\ell$ to perform linear algebra over $\Z/\ell\Z$.\\ \\
We next show that if for all $\ell$ dividing $L$, the $\Z/\ell\Z$ rank of $R_\ell$ is $|F|-1$, then the discrete logarithm of an element in the factorbase can be computed efficiently.
\begin{lemma}\label{gaussian} If for all $\ell$ dividing $L$, the $\Z/\ell\Z$ rank of the relation matrix $R_\ell$ is $|F|-1$, then we can compute a factorization $L = L_1L_2 \ldots L_i\ldots L_c$ into pairwise relatively prime factors such that modulo each factor $L_i$, through a sequence of row operations and row/column permutations $R$ can be efficiently written (with entries modulo $L_i$) in the form 
\[ \left( \begin{array}{cccccc}
r_{L_i}(1) & * & * &   \ldots  & * & *\\
0 & r_{L_i}(2) & * &\ldots   & * &  * \\
0 &  0 & r_{L_i}(3)  &\ldots  & *  &  * \\
\vdots &  \vdots & \vdots  &\ddots  & \vdots &  \vdots \\
0 & 0 &   0  &   \ldots & r_{L_i}(|F|-1) & *\\
0 & 0 & 0 &  \ldots  & 0 & 0\\
\vdots & \vdots & \vdots &  \ddots  & \vdots & \vdots\\
0 & 0 & 0 &  \ldots  & 0 & 0 \end{array} \right)\]\\
where $\forall j \in \{1,2,\ldots,|F|\}$, $r_{L_i}(j)$ is invertible modulo $L_i$. 
\end{lemma}
\noindent Denote by $r_{i,j}$ the entry in the $i^{th}$ row and the $j^{th}$ column of $R$. There exists an entry in $R$ that is not a multiple of $L$. We may assume that this entry is $r_{1,1}$ for otherwise we can permute the rows and columns appropriately.\\ \\
If $r_{1,1}$ is not invertible modulo $L$, then we have found $\gcd(r_{1,1},L)$, a non trivial factor of $L$.  We may extract the largest factor $\hat{L}$ of $L$ supported by the primes dividing $\gcd(r_{1,1},L)$ as follows. Set $M_1 := \gcd(r_{1,1},L)$, $M_2:=\gcd(r_{1,1},L/M_1)$, $M_3:=\gcd(r_{1,1},L/(M_1M_2))$ and so on until $M_i = 1$. Then $\hat{L} = L / (M_1M_2\ldots M_i)$. We recursively compute the desired matrix decomposition modulo $\hat{L}$ and modulo $L/\hat{L}$.\\ \\
If $r_{1,1}$ is invertible modulo $L$, then we may use it as a pivot and through row operations and make every other entry in the first row zero and the resulting submatrix with the first row and column removed is of $\Z/\ell\Z$ rank $|F|-2$ for every $\ell$ dividing $L$ and is dealt with recursively.\\ \\
Since at each step, we either reduce the number of columns by $1$ or reduce into two subproblems each with modulus at most half of $L$, the number of recursive steps in our algorithm is bounded by a polynomial in $\log(L)$ and $|F|$.\\ \\
Consider when we have reduced the number of columns to $1$ (say modulo a factor $L_i$ of $L$) by performing a sequence of row operations and row/column permutations. By applying the same sequence of operations on the all one vector, we have a system of relations in $(\Z^{|F|}/\Gamma_R)/L_i(\Z^{|F|}/\Gamma_R)$ of the form 
\[ \left( \begin{array}{cccccc}
r_{L_i}(1) & * & * &   \ldots  & * & *\\
0 & r_{L_i}(2) & * &\ldots   & * &  * \\
0 &  0 & r_{L_i}(3)  &\ldots  & *  &  * \\
\vdots &  \vdots & \vdots  &\ddots  & \vdots &  \vdots \\
0 & 0 &   0  &   \ldots & r_{L_i}(|F|-1) & *\\
0 & 0 & 0 &  \ldots  & 0 & x_{L_i}(|F|)  \\
\vdots & \vdots & \vdots &  \ddots  & \vdots & \vdots\\
0 & 0 & 0 &  \ldots  & 0 & x_{L_i}(|R|) \end{array} \right)     
\left( \begin{array}{c}
\alpha(L_i)_1\\
\alpha(L_i)_2\\
\alpha(L_i)_3\\
\vdots\\
\alpha(L_i)_{|F|-1}\\
\alpha(L_i)_{|F|}\\
\end{array} \right) =
\left( \begin{array}{c}
0\\
0\\
0\\
\vdots\\
0\\
0\\
\end{array} \right)\]\\
where $\forall j \in \{1,2,\ldots,|F|\}$, $r_{L_i}(j)$ is invertible modulo $L_i$ and $\alpha(L_i)_j \in (\Z^{|F|}/\Gamma_R)/L_i(\Z^{|F|}/\Gamma_R)$.\\ \\
This implies that $\alpha(L_i)_{|F|}$ generates $(\Z^{|F|}/\Gamma_R)/L_i(\Z^{|F|}/\Gamma_R)$.\\ \\
For a prime $\ell$ dividing $L_i$, since $\ell$ does not divide $L/L_i$, the largest power of $\ell$ dividing $L$ is the largest power of $\ell$ dividing $L_i$. Thus $\F_h^\times/(\F_h^\times)^{L_i} \cong \F_h^\times[L_i]$ where $\F_h^\times[L_i]$ denotes the $L_i$ torsion of $\F_h^\times$.\\ \\
Under the surjection $$\phi_{L_i}: (\Z^{|F|}/\Gamma_R)/L_i(\Z^{|F|}/\Gamma_R) \twoheadrightarrow \F_h^\times/(\F_h^\times)^{L_i} \cong \F_h^\times[L_i]$$
$\alpha(L_i)_{|F|}$ maps to a generator of $\F_h^\times[L_i]$.\\ \\
Since $|F_h^\times[L_i]| = L_i$, $L_i$ divides the order of $\alpha_{|F|}$. Thus $\forall j \in \{|F|,|F|+1,\ldots,|R|\}$,  since $x_{L_i}(j) \alpha(L_i)_{|F|} = 0$ it follows that $x_{L_i}(j)= 0 \mod L_i$.$\square$.
\begin{theorem}\label{dlog_gaussian}
If for all $\ell$ dividing $L$, the $\Z/\ell\Z$ rank of the relation matrix $R_\ell$ is $|F|-1$, then
\begin{enumerate} 
\item we can efficiently find a generator $\alpha_L$ for $\F_h^\times[L]$.
\item we can efficiently find the discrete logarithm of the image in $\F_h^\times[L]$ of an element in the factorbase with respect to $\alpha_L$.
\end{enumerate}
\end{theorem}
\noindent Consider the factorization $L = L_1 L_2 \ldots L_i \ldots L_c$ resulting from lemma \ref{gaussian}. From the proof of lemma \ref{gaussian}, for each $L_i$ in the factorization, we can find a generator $\alpha(L_i)_{|F|}$ of $(\Z^{|F|}/\Gamma_R)/L_i(\Z^{|F|}/\Gamma_R)$. A generator of  $\F_h^\times[L_i]$ can be found as $\phi_{L_i}(\alpha(L_i)_{|F|})$. Since the $L_i$ are pairwise relatively prime $$ \F_h^\times[L] = \prod_{i}\F_h^\times[L_i]$$ and we can extract a generator $\alpha(L_i)$ of $\F_h^\times[L]$ by the chinese remainder theorem. \\ \\
Given an element in $F_h$, from the proof of lemma \ref{gaussian}, we can perform a sequence of operations on the unit vector supported at that element to project its class $\beta$ in $(\Z^{|F|}/\Gamma_R)/L_i(\Z^{|F|}/\Gamma_R)$ as a power of $\alpha(L_i)_{|F|}$ thereby determining the discrete logarithm $\log_{\alpha(L_i)_{|F|}}(\beta)$ in $(\Z^{|F|}/\Gamma_R)/L_i(\Z^{|F|}/\Gamma_R)$. The discrete logarithm of the image of that factorbase element in $\F_h^\times/(\F_h^\times)^L \cong \F_h^\times[L_i]$ with respect to the image of $\phi_{L_i}(\alpha(L_i)_{|F|})$ is $\log_{\alpha(L_i)_{|F|}}(\beta)$.\\ \\
By chinese remainder theorem, we can thus determine the discrete logarithm of the image of a factorbase element in $\F_h^\times[L]$ with respect to the generator $\alpha(L_i)$. $\square$.\\ \\
With straightforward modifications, the algorithms developed in this section to prove lemma \ref{gaussian} and theorem \ref{dlog_gaussian} apply to the descent phase as well and lead to theorem \ref{descent_theorem}.
\section{Joux's Relation Generation Algorithm}\label{joux}
\subsection{Embedding}
The algorithm first proceeds by embedding $\F_{p^n}$ into an extension $\F_{q^{2m}}$ where $q$ is a power of $p$ such that $n \leq q$ and $m$ is a multiple of $n$ such that $q/2 < m \leq q$. In particular, we set $q := p^{\lceil \log_p(n)\rceil}$ and $m$ is chosen as the largest integral multiple of $n$ satisfying $q/2 < m \leq q$. We remark that our choice of embedding field $\F_{q^{2m}}$ is in certain cases larger than the one chosen in Joux's algorithm \cite{joux}.\\ \\
The field $\F_{q^{2m}}$ is constructed by adjoining a root $\zeta$ of an irreducible polynomial $g(x) \in \F_{q^2}[x]$ to $\F_{q^2}$ chosen in \S~\ref{poly_search}.\\ \\
The fact that we work over a specially chosen representation of the finite field which may differ from the input representation wherein the discrete logarithm is to be solved is not a concern for the following reason. We assume an explicit representation of $\F_{p^n}$ (see \cite{lenstra}) as an input. That is, a representation of $\F_{p^n}$ as an $\F_p$ vector space with a basis that allows efficient multiplication. For instance, regarding $\F_{p^n}$ as $\F_p[\mu]$ where $\mu$ is a root of a known irreducible degree $n$ polynomial is an explicit representation. Due to Lenstra \cite{lenstra}[Thm 1.2], an isomorphism between two explicit representations of a field of size $p^n$ can be computed deterministically in time polynomial in $n$ and $\log(p)$.
\subsection{Polynomial Search}\label{poly_search}
Let $C$ be a positive integer. We say that an integer is $q^{2C}$-smooth if and only if all its prime factors are at most $q^{2C}$.\\ \\
We define a polynomial $f(x) \in \F_{q^2}[x]$ to be ``good" if and only if the following four conditions are satisfied.
\begin{enumerate}
\item $f(x)$ has an irreducible factor of degree $m$ (call it $g(x)$).
\item The square of $g(x)$ does not divide $f(x)$.
\item $f(x)$ does not have linear factors.
\item $\gcd(\left|\F_{h/g}^\times \right|, q^{2m}-1)$ is $q^{2C}$-smooth.
\end{enumerate}
We set a degree bound $D$ and investigate the existence of $h_0(x),h_1(x) \in \F_{q^{2}}[x]$ each of degree bounded by $D$ such that $h(x) = h_1(x) x^q-h_0(x)$ is ``good". \\ \\
The existence of ``good" polynomials of the above form requires that $q+D$ is at least $m+2$  for otherwise we are left with a linear factor. To this end, if $m=q$, we assume $D>1$ and if $m=q-1$, we assume $D>0$.\\ \\
For $m>2$ and $r \geq m$, let $N_q(r,m)$ denote the number of polynomials in $\F_{q^2}[x]$ of degree $r \geq m$ that satisfy the first three conditions of being ``good" and let $P_q(r,m)= \frac{N_q(r,m)}{q^{2r}}$ denote the probability that a random polynomial of degree $r$ satisfies the first three conditions of being ``good". Let $s$ and $t$ be non negative integers such that $q+D-m = s(m-1)  + t$, where $t < m-1$. For a positive integer $k$, let $I_k$ denote the number of monic irreducible polynomials in $\F_{q^2}[x]$ of degree $k$.\\ \\
If $t \neq 1$, then $$N_q(q+D) \geq I_m \binom{I_{m-1}}{s} I_t$$
since we can chose an irreducible polynomial of degree $m$, $s$ irreducible polynomials of degree $m-1$ and one irreducible polynomial of degree $t$ and take their product to get a polynomial of degree $q+D$.
By substituting the lower bound $$I_k \geq \frac{q^k}{k} - \frac{q(q^{k/2}-1)}{(q-1)k}$$ in the above expression we get
$$P_q(q+D,m) = \frac{N_q(q+D,m)}{q^{2(q+D)}} \geq \frac{1}{m (m-1)^s t s! } \left( 1 - \mathcal{O}\left(\frac{1}{q^{t}}\right)\right).$$\\
Likewise, when $t=1$, it follows that $s \geq 1$ and we obtain $$N_q(q+D,m) \geq I_m \binom{I_{m-1}}{s-1} I_{m-2} I_{t+1}$$ $$\Rightarrow P_q(q+D,m) \geq \frac{1}{m (m-1)^{s-1} (m-2) (t+1) (s-1)! } \left( 1 - \mathcal{O}\left(\frac{1}{q^{t+1}}\right)\right).$$\\ \\
If we were to assume that a random polynomial of the form $h_1(x)x^q - h_0(x)$, where $h_0(x)$ and $h_1(x)$ are of degree at most $D$ satisfies the first three conditions of being ``good" with probability $P_q(q+D,n)$, then since $s = \mathcal{O}(D/m)$ choosing $$D = \Theta(\log_{q^2}(m (m-1)^s t s!)) = \Theta(1)$$ is sufficient to ensure the existence of $h_0(x)$ and $h_1(x)$ such that $h(x)$ is square free, has a degree $m$ factor and no linear factors.\\ \\
Heuristically it is likely that a large fraction of polynomials that satisfy the first three constraints also satisfy the fourth constraint on being ``good". \\ \\
For a polynomial that satisfies the first three conditions, if each of its factors excluding its degree $m$ factor is either of degree prime to $m$ or of degree bounded by $C$, then it is likely to satisfy the fourth condition.\\ \\
Consider positive integers $m^\prime, s^\prime$ and $t^\prime$ such that $m^\prime > m/2$, $t^\prime > 1$,  $q+D-m = s^\prime m^\prime + t^\prime$, $\gcd(m^\prime,m)=1$ and either $\gcd(t^\prime,m)=1$ or $t < C$. For such a choice, $\gcd(q^{2m^\prime}-1,q^{2m}-1)$ and $\gcd(q^{2t^\prime}-1,q^{2m}-1)$ are both likely to be $q^{\mathcal{O}(1)}$-smooth. Hence by taking an irreducible polynomial of degree $m$, $s^\prime$ irreducible polynomials of degree $m^\prime$ and an irreducible polynomial of degree $t^\prime$, we can construct a ``good" polynomial. From an analysis similar to the above computation of $P_q(q+D,m)$, we can conclude heuristically that choosing $D = \Theta(1)$ and $C =\Theta(1)$ are sufficient to guarantee the existence of the ``good" polynomials that we seek.
\begin{heuristic}\label{heu1} There exists positive integers $D,C$ such that for all prime powers $q$ and for all positive integers $2 < m \leq q$, there exists $h_0(x),h_1(x) \in  \F_{q^2}[x]$ of degree bounded by $D$ such that $h_1(x)x^q-h_0(x)$ is ``good".
\end{heuristic}
\textbf{Search for $h_0(x), h_1(x)$ and $g(x)$: }\textit{Fix constants $C,D$. Enumerate candidates for $h_0(x), h_1(x) \in \F_{q^2}[x]$ with each of their degrees bounded by $D$. For each candidate pair $(h_0(x) , h_1(x))$, factor $h(x) = h_1(x)x^q-h_0(x)$. If $h(x)$ is ``good", output $h_0(x), h_1(x)$ and the factor of degree $m$ and stop. If no such candidates are found, declare failure.}\\ \\
The search algorithm terminates after considering at most $q^{2(D+1)} = q^{\mathcal{O}(1)}$ candidate pairs. Factoring each candidate $h_1(x)x^q-h_0(x)$ takes time polynomial in the degree $q+D$ and $p$ using Berlekamp's deterministic polynomial factorization algorithm \cite{ber}. All four conditions of being good can be checked efficiently given the degrees of the irreducible factors and the corresponding powers in the factorization of $h(x)$. Thus, the search for $h_0(x), h_1(x)$ and hence $g(x)$ of the desired takes at most $q^{\mathcal{O}(1)}$ time.
\subsection{The Factorbase: A Small Generating Set} \label{gen}
Following Joux \cite{joux}, we choose a small subset $S \subseteq \F_{q^2}[\zeta]$ that generates $\F_{q^{2}}[\zeta]^\times$. F.R.K Chung proved that for all prime powers $s$, for all positive integers $r$ such that $(r - 1)^2<s$, for all $\mu$ such that $\F_{s^r} = \F_{s}[\mu]$, the set $\F_s + \mu$ generates $\F_{s^r}^\times$ \cite[Thm. 8]{chung}\cite[Ques 1.1]{wan}. Since $m \leq q$, setting $S :=  \F_{q^2} + \zeta $ ensures that the subgroup generated by $S$, $\langle S \rangle = \F_{q^{2}}[\zeta]^{\times}$.\\ \\
Given that $\langle S \rangle = \F_{q^{2m}}^\times$, the next step is to determine the relations satisfied by the elements in $S$ so that we can determine $\F_{q^{2}}[\zeta]$ as the free abelian group generated by $S$ modulo the relations.\\ \\
If $h(x)$ were to have a linear factor, then the relation generation step will not relate that linear factor to the rest of the linear polynomials in the factor base. As a result, we would have to exclude that linear factor from the factor base and F.R.K Chung's theorem that ensures  $\Z^{|F|}/\Gamma_g \cong \F_g^\times$ would no longer apply. It is to circumvent this that we insisted that $h(x)$ have no linear factors.\\ \\
For a technical reason, $S$ is first extended to the set $F := h_1(\zeta) \cup \{\lambda\} \cup S$, where $\langle \lambda \rangle = \F_{q^2}^\times$. We will call $F$ as the factorbase. An identity in $\langle F \rangle \cong \F_{q^{2m}}^\times$ of the form $\prod_{\beta \in F} \beta^{e_\beta} = 1$ for integers $e_\beta$ is called as a relation and it can be identified with the relation vector $(e_\beta, \beta \in F) \in \Z^{|F|}$ indexed by elements in $F$.
\subsection{Joux's Relation Generation Algorithm}\label{relation}
The relation search step begins with the following identity over $\F_{q^2}[x]$ $$\prod_{\alpha \in \F_{q}}{x-\alpha} = x^q - x.$$
For \[\mathfrak{m} = \left( \begin{array}{cc}
a & b  \\
c & d   \end{array} \right) \in GL(2,q^2), \]
the substitution $x \mapsto \frac{a\zeta+b}{c\zeta+d}$ yields\\ $$ \prod_{\alpha \in \F_q} \frac{(a-\alpha c)\zeta + (b-\alpha d)}{(c \zeta +d)^{q}} = \frac{(c \zeta + d)(a \zeta + b)^q - (a\zeta+b)(c\zeta+d)^q }{(c\zeta+d)^{q+1}}$$  
$$\Rightarrow (c\zeta+d) \prod_{\alpha \in \F_q} ((a-\alpha c)\zeta + (b-\alpha d)) = (c \zeta + d)(a \zeta + b)^q - (a\zeta+b)(c\zeta+d)^q. $$
Linearity of raising to the $q^{th}$ power implies $$ (c\zeta+d) \prod_{\alpha \in \F_q} ((a-\alpha c)\zeta + (b-\alpha d)) = (c \zeta + d)(a^q \zeta^q + b^q) - (a \zeta+b)(c^q \zeta^q+d^q).$$
By substituting $\zeta^q = \frac{h_0(\zeta)}{h_1(\zeta)}$, the right hand side becomes $$\frac{(ca^q-ac^q) \zeta h_0(\zeta) + (da^q-bc^q) h_0(\zeta) + (cb^q - ad^q) \zeta h_1(\zeta) + (db^q-bd^q) h_1(\zeta)}{h_1(\zeta)}.$$
Consider the numerator of the above expression as the polynomial $$N_{\mathfrak{m}}(x) : =  (ca^q-ac^q) x h_0(x) + (da^q-bc^q) h_0(x) + (cb^q - ad^q) x h_1(x) + (db^q-bd^q) h_1(x) $$ evaluated at $\zeta$. The degree of $N_{\mathfrak{m}}(x)$ is bounded by $D+1$. If $N_{\mathfrak{m}}(x)$ factors in to linear factors over $\F_{q^2}[x]$, then we get the following relation in $\langle F \rangle$ $$(c\zeta+d) h_1(\zeta) \prod_{\alpha \in \F_q} ((a-\alpha c)\zeta + (b-\alpha d))  = N_{\mathfrak{m}}(\zeta).$$
The above expression can be written as a product of an element $\mu \in \F_{q^2}^\times$ times $h_1(\zeta)$ times a fraction of products of monic linear polynomials in $\zeta$ over $\F_{q^2}$ being equal to $1$. By expressing the element $\mu$ in $\F_{q^2}^\times$ as a power of $\lambda$ by computing a discrete logarithm over $\F_{q^2}^\times$, we indeed get a relation in $\langle F \rangle$.\\ \\
The reason for choosing to work over $\F_{q^2}$ instead of $\F_q$ is that for every choice of $a,b,c,d \in \F_q$, the relation it yields becomes $\zeta^q-\zeta = \prod_{\alpha \in \F_q} (\zeta-\alpha)$. Thus, we have to work over an extension of $\F_q$ where the $q^{th}$ power map would be non trivial and $\F_{q^2}$ is the smallest such extension.\\ \\
For an $e \in \F_{q^2}^\times$, the substitutions $x \mapsto \frac{a \zeta + b}{c \zeta + d}$ and $x \mapsto \frac{ae \zeta + be}{ce \zeta + de}$ are identical and will lead to the same relation. Thus, the possible choices for $a,b,c,d \in \F_{q^2}$, that could lead to distinct relations can at best be identified with elements in $PGL(2,q^2)$.\\ \\ 
Further, every element in $PGL(2,\F_q)$ gives rise to the same identity $\prod_{\alpha \in \F_{q}}{\zeta-\alpha} = \zeta^q - \zeta$. More generally, every element in the left coset of $PGL(2,q)$ in $PGL(2,q^2)$ yields the same relation \cite{bgjt}. Thus the possible choices for $\mathfrak{m}$ can be identified with a set of representatives $\mathcal{P}_q$ of the left cosets of $PGL(2,q)$ in $PGL(2,q^2)$. The cardinality of $\mathcal{P}_q$ is $q(q^2+1) = \Theta(q^3)$.\\ \\
\textbf{Relation Generation:} \textit{For every $\mathfrak{m} \in \mathcal{P}_q$, compute the numerator $N_{\mathfrak{m}}(x)$ and if it factors into linear factors over $\F_{q^2}[x]$, add the relation obtained as a row to the relation matrix $R$.\\ \\
Add the relation corresponding to the identity $\lambda^{q^2-1} = 1$ to $R$.}\\ \\
The relation generation step can be performed in $q^{\mathcal{O}(1)}$ time since the set of representatives $\mathcal{P}_q$ can be constructed in $q^{\mathcal{O}(1)}$ time and factoring the numerator polynomial using Berlekamp's deterministic factoring algorithm takes $q^{\mathcal{O}(1)}$ time as the numerator polynomial is of constant degree. We have to express the constant $\F_{q^2}^\times$ factor in the relation as a power of $\lambda$, but that can be accomplished by solving the discrete logarithm in $\F_{q^2}^\times$ exhaustively in $\mathcal{O}(q^2)$ time. 
\subsection{Relation Generation Heuristic}
In this subsection, we argue under a heuristic assumption that the relation generation algorithm does indeed produce enough relations to solve the discrete logarithm problem between elements expressed as products in the factorbase.\\ \\
The probability that a random polynomial of degree at most $D+1$ factors into linear factors is roughly $\frac{1}{(D+1)!}$ \cite{pgf}. If the numerator polynomials $N_{\mathfrak{m}}(x)$ that appear in the relation generation phase behave as random polynomials of the same degree with respect to their probability of splitting in to linear polynomials, then the expected number of trials required to get a relation is $(D+1)!$.  Since $D$ is a constant independent of $q$ and $n$, the expected number of rows of $R$ is a constant fraction of $\Theta(q^3)$.\\ \\
Since the dimension of the lattice $|F|$ is at most  $q^2+2$ and $\Gamma_R$ is the lattice generated by $\Theta(q^3)$ points, it is overwhelmingly likely that $\Gamma_R = \Gamma_h$, which makes the weaker claim of the heuristic \ref{relation_lattice} below even more plausible.
\begin{heuristic}\label{relation_lattice}
The generated relation lattice $\Gamma_R$ is large enough to ensure that the greatest common divisor of $q^{2m}-1$ and the cardinality of the second largest invariant factor of $Z^{|F|}/\Gamma_R$ is $q^{2C}$-smooth.
\end{heuristic}
\noindent By applying theorem \ref{discrete_log_theorem} to the relation lattice obtained by Joux's algorithm with modified polynomial selection, we have theorem \ref{dlog_heuristic_theorem}. 
\begin{theorem}\label{dlog_heuristic_theorem}
If heuristic \ref{relation_lattice} is true, then \\ \\
(1) a generator $\mu$ of $\F_g^\times$ can be found in $q^{\mathcal{O}(1)}$-time.\\ \\
(2) $\forall \beta \in F$, a $\theta_\beta \in \Z$ such that $\mu^{\theta_\beta} = \beta$ can be found the in $q^{\mathcal{O}(1)}$-time. 
\end{theorem}
\section{The Barbulescu-Gaudry-Joux-Thome Descent}\label{descent}
\ \\ Given two elements $\gamma,\eta \in \F_g^\times$, we are interested in deciding if $\gamma \in \langle \eta \rangle$ and if so finding an integer $\log_\eta(\gamma)$ (determined modulo the order of $\eta$) such that $\gamma = \eta ^{\log_\eta(\gamma)}$. This is a generalization of the discrete logarithm problem.\\ \\
From \S~\ref{discrete_log_factorbase}, given two elements in $\F_g^\times$, each expressed as a product of powers of elements in the factorbase $F$, we can efficiently decide if the first element is in the subgroup generated by the other and if so compute the discrete logarithm of the first element with respect to the second element as the base.\\ \\
The descent step takes an arbitrary element in $\F_g^\times$ and attempts to express it as a product of powers of elements in the factorbase. Thus if the descent succeeds in expressing $\gamma$ and $\eta$ as products of powers of elements in the factorbase, then we would have solved the discrete logarithm problem.\\ \\
In a recent breakthrough, Barbulescu, Gaudry, Joux and Thome proposed a descent algorithm that under certain heuristic assumptions succeeds in expressing an element in $\F_g^\times$ as a product over the factorbase in 
quasi polynomial time \cite{bgjt}.\\ \\
An issue was identified in \cite{cwz} as a potential trap that prevents the descent from succeeding. Further a trap avoiding version of the descent was proposed in \cite{cwz} wherein certain relations generated are excluded. We propose a modification to the descent in \cite{bgjt} wherein some of the relations excluded in fear of traps can be salvaged. Further, the salvaged relations aid in further breaking the symmetry between $g(x)$ and the other factors of $h(x)$. A brief account of the descent algorithm in \cite{bgjt} and our proposed modification follows.\\ \\
An element $\eta \in \F_g^\times$ is presented to the descent algorithm as a polynomial $P(x) \in \F_q^2[x]$ of degree $w$ such that $\eta = P(\zeta)$ and $w <m$.\\ \\
We may assume that $P(x)$ and $h(x)$ do not share a non constant factor. Otherwise, raise $P(x)$ to a random power, then divide by $h(x)$ and call the remainder $P^\prime(x)$. It is likely that $P^\prime(x)$ and $h(x)$ do not share a factor and hence we can start the descent from $P^\prime(x)$.\\ \\
The first step in the descent attempts to reduce the problem to performing a descent on a set of inputs each of degree $w/2$ or less. To this end, a set of multiplicative relations modulo $h(x)$ relating the $\F_{q^2}$ translates of $P(x)$ with polynomials of degree at most $w/2$ are obtained. From the relations obtained, we then attempt to express modulo $h(x)$ each $\F_{q^2}$ translate of $P(x)$ as a product of powers of polynomials of degree at most $w/2$ and powers of $\lambda$ and $h_1(x)$.\\ \\
The first step again starts with the identity $$\prod_{\alpha \in \F_{q}}{x-\alpha} = x^q - x.$$
For \[\mathfrak{m} = \left( \begin{array}{cc}
a & b  \\
c & d   \end{array} \right) \in \mathcal{P}_q, \]
the substitution $x \mapsto \frac{aP(x)+b}{cP(x)+d}$ yields\\ $$ \prod_{\alpha \in \F_q} \frac{(a-\alpha c)P(x) + (b-\alpha d)}{(c P(x) +d)^{q}} = \frac{(c P(x) + d)(a P(x) + b)^q - (aP(x)+b)(cP(x)+d)^q }{(cP(x)+d)^{q+1}}$$  
$$\Rightarrow (cP(x)+d) \prod_{\alpha \in \F_q} ((a-\alpha c)P(x) + (b-\alpha d)) = (c P(x) + d)(a P(x) + b)^q - (aP(x)+b)(cP(x)+d)^q. $$
Linearity of raising to the $q^{th}$ power implies $$ (cP(x)+d) \prod_{\alpha \in \F_q} ((a-\alpha c)P(x) + (b-\alpha d)) = (c P(x) + d)(a^q \tilde{P}(x^q) + b^q) - (a P(x)+b)(c^q \tilde{P}(x^q)+d^q).$$ where $\tilde{P}(x)$ is $P(x)$ with its coefficients raised to the $q^{th}$ power.\\ \\
By substituting $x^q = \frac{h_0(x)}{h_1(x)}$, we obtain a congruence module $h(x)$. Under the substitution, the right hand side becomes $$(c P(x) + d)(a^q \tilde{P}\left(\frac{h_0(x)}{h_1(x)}\right) + b^q) - (a P(x)+b)(c^q \tilde{P}\left(\frac{h_0(x)}{h_1(x)}\right)+d^q) $$ which can be expressed as a fraction $$N_{\mathfrak{m},P}(x)/D_{\mathfrak{m},P}(x)$$ where $N_{\mathfrak{m},P}(x) \in \F_q^2[x]$ is of degree bounded by $(1+D)w$ and $D_{\mathfrak{m},P}(x) \in \F_q^2[x]$ is a power of $h_1(x).$\\ \\
If $N_{\mathfrak{m},P}(x)$ were to factor over $\F_{q^2}[x]$ into a product of irreducible factors each of degree bounded by $w/2$, then we obtain a relation of the form $$ \prod_{\beta \in \F_{q^2}} (P(x)-\beta)^{e_\beta}  =  \lambda^{b_\lambda}h_1(x)^{e_{h_1}} \prod_{u \in U_{\mathfrak{m},P}} u(x)^{b_u} \mod h(x)$$
where $\forall \beta \in \F_{q^2}, e_\beta \in \{0,1\}$ (See \cite{bgjt} for a proof), $U_{\mathfrak{m},P}$ denotes a set of monic irreducible polynomials in $\F_{q^2}[x]$ of degree bounded by $w/2$, $b_\lambda,b_{h_1} \in \Z$ and $\forall u \in U_{\mathfrak{m},P},\ b_u \in \Z-\{0\}$.\\ \\
Let $U_P$ denote the union of the sets $U_{\mathfrak{m},P}$ as $\mathfrak{m}$ ranges over elements in $\mathcal{P}_q$ that result in a relation. If sufficiently many relations are generated, then we can express every $\F_{q^2}$ translate of $P(x)$ as a product over powers of polynomials in $U_P$ and powers of $h_1$ and $\lambda$. We recursively perform the descent on the elements in $U_P$ until we decompose into linear factors.\\ \\ 
In \cite{cwz}, the following scenario was identified as a possible trap that prevents a descent step from working. Consider a $\mathfrak{m} \in \mathcal{P}_q$ that results in the following relation
\begin{equation} \prod_{\beta \in \F_{q^2}} (P(x)-\beta)^{e_\beta}  =  \lambda^{b_\lambda}h_1(x)^{e_{h_1}} \prod_{u \in U_{\mathfrak{m},P}} u(x)^{b_u} \mod h(x) \end{equation}
where in a $v(x) \in U_{\mathfrak{m},P}$ appears such that $v(x)$ divides $h(x)$. In the next step, one tries to relate $v(x)$ and its $\F_{q^2}$ translates modulo $h(x)$ to powers of irreducible polynomials of degree at most half of $\deg(v)$. However, since $v(x)$ is irreducible in $\F_{q^2}[x]$ and not a unit modulo $h(x)$, $v(x)$ would never appear in a relation in $\F_h^\times$ involving only the $\F_q^2$ translates of $v(x)$ and smaller degree polynomials. The trick of raising $v(x)$ to a random power modulo $h(x)$ is not available \footnote{In \cite{bgjt2}[Prop 10], a descent step starting from $v(x)$ for the case when $D \leq 2$ is described.} in the intermediate steps since it might raise the degree.\\ \\
To remedy this scenario, it was proposed in \cite{cwz} to not use relations where in such a $v(x)$ is involved. As a result the necessity to perform a descent on $v(x)$ would not arise. This trap avoidance strategy comes at a cost since certain relations are not utilized.\\ \\
While it is true that the image of $v(x)$ is not a unit in $\F_h$, it is a unit in $\F_g$ which is the field we are ultimately interested in. Further, since the number of factors of $h(x)$ is small compared to the number of relations we expect to get, in addition to the $\F_{q^2}$ translates of $P(x)$, we can try to eliminate the factors of $h(x)$ that appear.\\ \\ 
The modification to the descent step we propose is that at each step we attempt to express every element in $$\left\{\frac{P(x)-\beta }{\gcd(P(x)-\beta,h(x)/g(x))} | \beta \in \F_{q^2}\right\} \bigcup G_P$$ modulo $g(x)$ as a product of powers of polynomials of degree at most $w/2$ and powers of $h_1(x)$ and $\lambda$. Here $G_P$ is the set of all factors of $h(x)/g(x)$ that appear in the descent step involving $P(x)$. A formal definition of $G_P$ is in the description below.\\ \\
Say $N_{\mathfrak{m},P}$ does factor over $\F_{q^2}[x]$ into a product of irreducible factors each of which is either of degree bounded by $w/2$ or a factor of $h(x)$.\\ \\ 
The image of every factor of $h(x)/g(x)$ in $\F_g$ is a unit and hence can be inverted resulting in a relation of the form
$$\left( \prod_{i=1}^k g_i(x)^{s_{\mathfrak{m},i}}\right) \times \prod_{\beta \in \F_{q^2}} \left( \frac{P(x)-\beta}{\gcd(P(x)-\beta,h(x)/g(x))}\right)^{r_{\mathfrak{m},\beta}}   =  \lambda^{c_{\mathfrak{m},\lambda}} h_1(x)^{c_{\mathfrak{m},h_1}} \prod_{u \in V_{\mathfrak{m},P} } u(x)^{c_{\mathfrak{m},u}} \mod g(x)$$ where $V_{\mathfrak{m},P}$ is a set of monic irreducible polynomial of degree at most $w/2$ each of whose elements is not a factor of $h(x)$. Here $c_{\mathfrak{m},\lambda},c_{\mathfrak{m},h_1} \in \Z$ and $\forall i \in \{1,2,\ldots,k\}, s_{\mathfrak{m},i} \in \Z$ and $ \forall \beta\in \F_{q^2}, r_{\mathfrak{m},\beta} \in \Z $ and $\forall u \in V_{\mathfrak{m},P}, c_{\mathfrak{m},u} \in \Z-\{0\}$.\\ \\
For $i \in \{1,2,\ldots,k\}$, let $\mathcal{V}_i: \F_{q^2}(x) \longrightarrow \Z$ denote the valuation at $g_i(x)\F_{q^2}[x]$.\\ \\
If $\forall i \in \{1,2,\ldots,k\}$ and $\forall \beta \in \F_{q^2}$, $\mathcal{V}_i(P(x) - \beta) = 0$, then none of the factors of $h(x)$ can divide $N_{\mathfrak{m},P}$ and there is no need to look out for traps.\\ \\
If $\exists \beta \in \F_{q^2}$ and $\exists i \in \{1,2,\ldots,k\}$ such that $\mathcal{V}_i(P(x)-\beta) > 1$, then every $\mathfrak{m}$ that results in a relation involving $P(x)-\beta$ satisfies $\mathcal{V}_i(N_{\mathfrak{m},P})>1$. If $$\mathcal{V}_i(N_{\mathfrak{m},P}) = \sum_{\beta \in \F_{q^2}}\mathcal{V}_i(P(x)-\beta),$$ we can cancel the powers of $g_i(x) \mod g(x)$ and end up with $s_{\mathfrak{m},i}=0$. Else, the cancellation will result in $s_{\mathfrak{m},i} \neq 0$.\\ \\
Define $G_P$ to be $\left\{g_i(x)\ |\ \exists \mathfrak{m} \in \mathcal{P}_q: s_{\mathfrak{m},i} \neq 0\right\}$. In particular, $G_P$ is a subset of the set of irreducible factors of $h(x)$ that divide a translate of $P(x)$.\\ \\
If $N_{\mathfrak{m},P}(x)$ does factor over $\F_{q^2}[x]$ into a product of irreducible factors each of which is either of degree bounded by $w/2$ or divides $h(x)$, then from the relation obtained, form the relation vector $$R_{\mathfrak{m},P}:=\left(s_{\mathfrak{m},i}, r_{\mathfrak{m},\beta}\right)_{g_i(x) \in G_P, \beta \in \F_q^2} \in \Z^{|G_P|+q^2}$$ indexed by the elements in $G_P$ and $\F_{q^2}$. Let $M_P$ be the matrix consisting of $R_{\mathfrak{m},P}, \mathfrak{m} \in P_q$ as the rows where we only consider $\mathfrak{m}$ that resulted in a relation. Let $V_P$ denote the union of the sets $V_{\mathfrak{m},P}$ as $\mathfrak{m}$ ranges over elements in $\mathcal{P}_q$ that result in a relation.\\ \\
Recall that $L$ equals $q^{2m}-1$ divided by the largest $q^{2C}$-smooth factor of $q^{2m}-1$.\\ \\
For every prime $\ell$ dividing $L$, if $M_{P}$ is of rank $q^2+|G_P|$ over $\Z/\ell\Z$, then $\forall \beta \in \F_{q^2}$ and $\forall g_i(x) \in G_P$, we can express the projections of $\frac{P(x)-\beta}{\gcd(P(x)-\beta,h(x)/g(x))}$ and $g_i(x)$ in $\F_g^\times[L]$ as a product of powers of projections of polynomials of degree bounded by $w/2$, $\lambda$ and $h_1(x)$.\\ \\
The complement of $\F_g^\times[L]$ under direct product in $\F_g^\times$ is $q^{2C}$-smooth and can be accounted for using the Pohlig-Hellman algorithm.\\ \\
The degree of $N_{\mathfrak{m},P}(x)$ is bounded by $(1+D)w$. The probability that a random polynomial of degree at most $(1+D)w$ factors into irreducible factors of degree bounded by $w/2$ is around $1/(2(1+D))!$ which is a constant independent of $w$. Since there are $q(q^2+2)$ choices for $\mathfrak{m}$, if $N_{\mathfrak{m},P}(x)$ were to factor into irreducible polynomials of degree bounded by $w/2$ with a probability identical to that of a random polynomial of the same degree, then we expect to get at least $\Theta(q^3)$ relations. The number of columns in $M_P$ is bounded by $q^2+(q+D-m)/2$ and is likely to be close to $q^2$. The number of relations generated is likely to far exceed the number of columns in $M_P$ and thus $M_P$ is likely to have rank $q^2+|G_p|$ over $\mathbb{Q}$. Further, for every $\ell$ dividing $L$, since the $\ell$-primary part of $\F_h^\times$ is cyclic the concern raised in the introduction section do not arise and it is likely that the $\Z/\ell\Z$ rank of $M_P$ is $q^2+|G_P|$ and the following heuristic is plausible.
\begin{heuristic}\label{rank}{For all $P(x) \in \F_{q^2}[x]$ of degree less than $m$, if $P(x)$ is relatively prime to $h(x)$, then for every prime $\ell$ dividing $L$, $M_{P}$ is of rank $q^2+|G_P|$ over $\Z/\ell\Z$.}
\end{heuristic}
\noindent If heuristic \ref{rank} is true, then at each step of the descent, we reduce the problem of descent from a polynomial $P(x)$ of degree $w$ to the problem of descent from a set of polynomials $V_P$ of degree at most $w/2$. A step in the descent can be performed in $q^{\mathcal{O}(1)}$ time using a straight forward modification of the algorithms developed in \S~\ref{alternate} to prove lemma \ref{gaussian} and theorem \ref{dlog_gaussian}. Further, the size of $V_P$ is at most $\mathcal{O}(q^2w)$ \cite{bgjt}.  Since at each step we have at most $\mathcal{O}(q^2w)$ new descent steps involving polynomials of degree at most $w/2$, the total running time of the descent is $q^{\mathcal{O}(\log w)} = q^{\mathcal{O}(\log m)}$ and we have theorem \ref{descent_theorem}.
\begin{theorem}\label{descent_theorem}
If heuristic \ref{rank} is true, then in $q^{\mathcal{O}(\log m)}$ time we can find the discrete logarithm of an element in $\F_g^\times \cong \F_{q^{2m}}^\times$. 
\end{theorem}
\noindent Since $q^{\mathcal{O}(\log m)}$ is bounded by $(p n)^{\mathcal{O}(\log n)}$, we can find discrete logarithms in $\F_{p^n}^\times$ in $(p n)^{\mathcal{O}(\log n)}$ time and the algorithm is efficient in small characteristic.\\ \\
If Heuristic \ref{rank} fails for some $u(x)$ in the descent tree starting from a polynomial $P(x)$, then we may try again by taking a random power of $P(x)$ modulo $h(x)$.\\ \\ 

\bibliographystyle{amsplain}

\begin{thebibliography}{99}


\bibitem[Adl]{adl} L. M. Adleman. ``A subexponential algorithm for the discrete logarithm problem with applications to cryptography". In {\em Foundations of Computer Science, 1979., 20th Annual Symposium on}, pages 55--60. IEEE, 1979.

\bibitem[Adl1]{adl1} L. M. Adleman, ``The function field sieve", In {\em Algorithmic number theory-ANTS I}, volume 877 of {\em Lecture Notes in Computer Science}, pages 108--121. Springer, 1994.

\bibitem[AH]{ah} L. M. Adleman, M-D Huang, ``Function Field Sieve Method for Discrete Logarithms over Finite Fields", Information and Computation, Volume 151, Issues 1Ð2, 25 May 1999, Pages 5Ð16.

\bibitem[BGJT]{bgjt} R. Barbulescu, P. Gaudry, A. Joux , E. Thome, ``A quasi-polynomial algorithm for discrete logarithm in finite fields of small characteristic", http://arxiv.org/abs/1306.4244

\bibitem[BGJT2]{bgjt2} R. Barbulescu, P. Gaudry, A. Joux , E. Thome, ``A quasi-polynomial algorithm for discrete logarithm in finite fields of small characteristic", http://eprint.iacr.org/2013/400.pdf

\bibitem[Ber]{ber} E. R. Berlekamp, ``Factoring Polynomials Over Finite Fields", Bell System Technical Journal 46 (1967): 1853Ð1859.

\bibitem[CWZ]{cwz} Q. Cheng, D. Wan and J. Zhang, ``Traps to the BGJT-Algorithm for Discrete Logarithms" http://arxiv.org/abs/1310.5124

\bibitem[Cop]{cop} D. Coppersmith. Fast evaluation of logarithms in fields of characteristic two. {\em Information Theory, IEEE Transactions on}, 30(4):587--594, 1984.

\bibitem[Chu]{chung} F.R.K Chung, ``Diameters and Eigenvalues", J. Amer. Math. Soc. 2 (1989), no. 2, 187{196}.

\bibitem[DH]{dh} W. Diffie and M. Hellman. New directions in cryptography. {\em Information Theory, IEEE Transactions on}, 22(6):644--654, 1976.

\bibitem[ElG]{elg} T. ElGamal . A Public-Key Cryptosystem and a Signature Scheme Based on Discrete Logarithms.{\em IEEE Transactions on Information Theory},31 (4): 469Ð472.

\bibitem[GGMZ]{gggz} Faruk Gologlu, Robert Granger, Gary McGuire, and Jens Zumbragel, On the function field sieve and the impact of higher splitting probabilities: Application to discrete logarithms in $\mathbb{F}_{2^{1971}}$. Cryptology ePrint Archive, Report 2013/074, 2013.

\bibitem[GGMZ1]{gggz1} F. Gologlu, R. Granger, G. McGuire and J. Zumbragel,`` Discrete Logarithms in GF($2^{1971}$)", NMBRTHRY List, Feb 2013.

\bibitem[GGMZ2]{gggz2} F. Gologlu, R. Granger, G. McGuire and J. Zumbragel,`` Discrete Logarithms in GF($2^{6120}$)", NMBRTHRY List, Apr 2013.

\bibitem[Gor]{gor} Daniel M Gordon. Discrete logarithms in {{GF(p)}} using the number field sieve. {\em SIAM Journal on Discrete Mathematics}, 6(1):124--138, 1993.

\bibitem[HN]{primitive}  M-D Huang and A. K. Narayanan ``Finding primitive elements in finite Þelds of small characteristic", http://arxiv.org/pdf/1304.1206v4.pdf

\bibitem[JL]{jl} Antoine Joux and Reynald Lercier. The function field sieve in the medium prime case. In {\em Advances in Cryptology-EUROCRYPT 2006}, volume 4005 of {\em
 Lecture Notes in Computer Science}, pages 254--270. Springer, 2006.

\bibitem[JLSV]{jlsv} Antoine Joux, Reynald Lercier, Nigel Smart, and Frederik Vercauteren. The number field sieve in the medium prime case. In {\em Advances in Cryptology-CRYPTO 2006}, pages 326--344. Springer, 2006.

\bibitem[Jou]{joux} Antoine Joux. A new index calculus algorithm with complexity {{L}}$(1/4+o(1))$ in very small characteristic. Cryptology ePrint Archive, Report 2013/095, 2013.

\bibitem[Jou2]{joux2} A. Joux, `` Discrete Logarithms in GF($2^{4080}$)", NMBRTHRY List, March 2013.

\bibitem[Len]{lenstra} H.W Lenstra, ``Finding isomorphism between finite fields", Math. Comp., 56 (1991), pp. 329Ð347.

\bibitem[PGF]{pgf} D. Panario, X. Gourdon, P. Flajolet, ``An Analytic Approach to Smooth Polynomials over Finite Fields", ANTS 1998: 226-236

\bibitem[PH]{ph} S. Pohlig, M. Hellman,  ``An Improved Algorithm for Computing Logarithms over GF(p) and its Cryptographic Significance", IEEE Transactions on Information Theory (24): 106Ð110, (1978).

\bibitem[Wan]{wan} D. Wan, ``Generators and irreducible polynomials over finite fields", Math. Comp. 66 (219) (1997) 1195Ð1212.

\end{thebibliography}

\end{document}